%Paper: gr-qc/9506039
%From: Philip Tuckey <pht@mppmu.mpg.de>
%Date: Tue, 20 Jun 1995 14:38:17 +0200
%Date (revised): Sat, 29 Jul 1995 16:44:24 +0200
%Date (revised): Thu, 14 Sep 1995 18:01:06 +0200

% PLAIN TEX, USES HARVMAC.
% FOR LANLMAC: Comment out the \input harvmac line, the \footline statement
% on line 30, and the \def\footlab on line 34, and uncomment the
% \input lanlmac line and the two corresponding lanlmac definitions.

\input harvmac
%\input lanlmac                                                      %LANLMAC

% lanlmac options
%\draftmode
\sequentialequations

% Moved here from later in lanlmac
% Done in \Title in standard lanlmac format
\nopagenumbers % no number on title page
\pageno=0 % title page comes out on its own in reduced style
% Normally done correctly automatically (i.e. in \output)
\def\firsttextpageno{1} % Number of first page after title page

% Modifications to lanlmac
\baselineskip12pt
\footskip11pt
\parindent0pt
\parskip9pt
% Typeset title page at same size as others in reduced style
\pageno=1\hstitle=\hsbody\def\abstractfont{\tenpoint}
% In addition to previous, make title page be page 1 and number it
\def\firsttextpageno{2}
\footline={\hss\tenrm\folio\hss}
%\footline={\hss\tenrm\hyperdef\hypernoname{page}\folio\folio\hss}   %LANLMAC

% Macros
\def\footlab#1{\xdef #1{\the\ftno}\writedef{#1\leftbracket#1}\wrlabeL{#1=#1}}
\def\bra{\langle}
\def\C{\hbox{\bf C}}
\def\F{{\cal F}}
\def\hat{\widehat}
\def\hb{{\cal H}}
\def\ket{\rangle}
\def\lg{\hbox{\rm SO$_0$(3,1)}}
\def\mm{i}
\def\nn{j}
\def\one{{\bf 1}_\F}
\def\onex{{\bf 1}_x}
\def\pg{\hbox{\rm ISO$_0$(3,1)}}
\def\PM{PM}
\def\prug{Prugove\v{c}ki}
\def\R{\hbox{\bf R}}
\def\ra{R}
\def\so{s^{\mkern-8mu\circ}}
\def\tilde{\widetilde}
\def\ua{\underline{a}}
\def\ue{e}%\def\ue{\underline{e}}
\def\uk{\underline{K}}
\def\uomega{\omega}%\def\uomega{{\underline{\omega}}}
\def\uphi{\Phi}%\def\uphi{\underline{\Phi}}
\def\upsi{\Psi}%\def\upsi{\underline{\Psi}}
\def\uq{\underline{q}}
\def\us{s}%\def\us{\underline{s}}
\def\usig{\sigma}%\def\usig{\underline{\sigma}}
\def\utheta{\theta}%\def\utheta{\underline{\theta}}
\def\uu{u}%\def\uu{\underline{u}}
\def\uv{\underline{v}}

\def\V{{V}}

\def\W{{\cal W}}

% References
\lref\rdi{Drechsler W 1982 {\it Ann.~Inst.~Henri Poincar\'e\/ \bf 37}
155--184, 1984 {\it Fortschr.\ Phys.\/ \bf 32} 449--472}
\lref\re{Ehlers J 1971 in {\it General relativity and cosmology\/}
ed.~B.K.~Sachs (New York: Academic Press)}
\lref\rfulling{Fulling S A 1973 {\it Phys.~Rev.\/ \rm D\bf 7} 2850--2862}
\lref\rgraud{Graudenz D 1994 {\it On the space-time geometry of quantum
systems\/} preprint CERN-TH.7516/94}
\lref\rhw{Higuchi A and Wald R M 1995 {\it Phys.~Rev.\/ \rm D \bf 51}
544--561}
\lref\rk{Kobayashi S 1956 {\it Can.~J.~Math.\/ \bf 8} 145--156}
\lref\rkn{Kobayashi S and Nomizu K 1963 {\it Foundations of differential
geometry\/} vol.~1 (New York: Wiley)}
\lref\rnw{Newton T D and Wigner E P 1949 {\it Rev.~Mod.~Phys.\/ \bf 21}
400}
\lref\rpi{\prug\ E 1986 {\it Stochastic quantum mechanics and quantum
spacetime\/} (Dordrecht: Reidel)}
\lref\rpii{\prug\ E 1992 {\it Quantum geometry: A framework for quantum
general relativity\/} (Dordrecht: Kluwer Academic)}
\lref\rpnewb{\prug\ E 1995 {\it Principles of quantum general
relativity\/} (Singapore; River Edge, NJ: World Scientific)}
\lref\rpnewp{\prug\ E 1994 {\it Class.~Quantum Grav.\/ \bf 11} 1981--1994}
\lref\runr{Unruh W G 1976 {\it Phys.~Rev.\/ \rm D\bf 14} 870--892}
\lref\runrw{Unruh W G and Wald R M 1984 {\it Phys.~Rev.\/ \rm D\bf 29}
1047--1056}
\lref\rw{Wald R M 1993 {\it Phys.~Rev.\/ \rm D \bf 48} R2377--R2381}
%Applications of a new proposal for solving the ``problem of time'' to some
%simple quantum cosmological models\/} preprint gr-qc/9407038}
\lref\ryellow{Choquet-Bruhat Y, DeWitt-Morette C and Bleick M 1985 {\it
Analysis, manifolds and physics, revised edition} (North-Holland)}
\lref\rlast{\prug\ E 1995 {\it On quantum-geometric connections and propagators
in curved spacetime}, preprint University of Toronto, Dept. of Mathematics}

% Title page
\hsize=\hstitle\abstractfont
\rightline{{\vbox{\hbox{MPI-PhT/95-37}
\hbox{August 1995}}}}
\vskip10mm\titlefont
\centerline{On quantum and parallel transport}
\vskip8pt
\centerline{in a Hilbert bundle over spacetime}
\vskip8mm\abstractfont\authorfont
\centerline{W.~Drechsler\footnote{$^\dagger$}{wdd@dmumpiwh.bitnet}}
\vskip6pt
\centerline{Philip A.~Tuckey\footnote{$^\ddagger$}{Supported by a
fellowship from the Alexander von Humboldt Foundation
\hfil\break pht@mppmu.mpg.de}}
\vskip6mm\sl
\centerline{Max-Planck-Institut f\"ur Physik}
\centerline{F\"ohringer Ring 6}
\centerline{D-80805 M\"unchen}
\centerline{Germany}

\vfill
\rm
\centerline{Abstract}\smallskip
We study the Hilbert bundle description of stochastic quantum mechanics in
curved
spacetime developed by \prug, which gives a powerful new framework for
exploring the quantum mechanical propagation of states in curved spacetime.
We concentrate on the quantum transport law in the bundle, specifically on
the information which can be obtained from the flat space limit. We give a
detailed proof that quantum transport coincides with parallel transport in
the bundle in this limit, confirming statements of \prug. We furthermore
show that the quantum-geometric propagator in curved spacetime proposed by
\prug, yielding a Feynman path integral-like formula involving integrations
over intermediate phase space variables, is Poincar\'e gauge covariant
(i.e.$\!$ is gauge invariant
except for transformations at the endpoints of the path) provided the
integration measure is interpreted as a ``contact point measure'' in the
soldered stochastic phase space bundle raised over curved spacetime.

\Date{} % Ends title page, plus does some other stuff
\pageno=\firsttextpageno % Set number of the first text page

\newsec{Introduction}

In this paper we investigate a new approach to quantum mechanics and
quantum field theory in curved spacetime, developed by \prug, which differs
from the standard approach in two central ways. The first of these is that
it uses the (special-) relativistic ``stochastic'' quantum mechanics
developed by \prug, which is based on a system of covariance for the
Poincar\'e group based on positive operator valued measures of a phase
space representation, rather than a system of imprimitivity constructed
with projector valued measures \rpi\rpii\ (and references therein). The
second major difference comes in the extension to curved spacetime, which
uses a fibre bundle over spacetime whose standard fibre is the usual
flat-space Hilbert or Fock space of the system considered
\rpii\rpnewp\rpnewb\ (and references therein). This allows the definition
of a particle to be given in terms of the properties of the fibres above
the points in spacetime, independently of the curvature of spacetime and of
the acceleration of any individual observer, or of other global properties
of the spacetime base. The curvature of spacetime
plays its role in the propagation of states from the fibre above one point
to that above another.

In the current paper we are specifically concerned with this ``quantum
propagation'' in the bundle over spacetime, guided by the
information which can be gained from the flat-space limit. Reference
\rpnewb\ postulates a Feynman path integral-like formula for the ``quantum
propagator'' in curved spacetime. We show that this quantum-geometric
propagator defined in \rpnewb\ has the correct flat space limit and
is Poincar\'e gauge covariant in curved spacetime
(i.e.$\!$ is gauge invariant except for
transformations at the endpoints of the path). To obtain this result
we use the invariance of
the integration measure at the contact points of base space and fibre at
the intermediate phase space integration points involved.

Standard problems in relativistic quantum mechanics include the choice of
the Hilbert space, the definition of position and momentum operators, and
the construction of a position probability density. In the case of a
spinless particle in Minkowski space, the Hilbert space is conventionally
chosen to be the positive frequency solutions of the Klein-Gordon
equation. It is well known that the conserved current constructed from such
a solution cannot be used to define a position probability density for the
particle, because its ``time'' component is not positive definite. Newton
and Wigner \rnw\ constructed position and momentum operators, whose
spectral resolution may be used to define probability densities on any
constant-time hypersurface (in an inertial coordinate system). However, as
is also well known, the resulting position probability densities do not
obey relativistic causality. Recently Wald \rw\ has discussed the choice of
the Hilbert space for a particle in a curved spacetime and background
fields, emphasising its arbitrariness, and has given a generalisation of
the Newton-Wigner operators to such cases.
In quantum field theory in curved space time, there is an
observer-dependent ambiguity in the vacuum state, and a corresponding
ambiguity in the definition of a particle \rfulling. A celebrated
consequence of this is the Unruh effect
\runr, which predicts that an observer accelerating in Minkowski space will
see a thermal bath of particles. As this effect has yet to be observed
experimentally, there remains the possibility that it is an artifact of
inadequacies in the standard quantisation procedure.

In pioneering work, \prug\ has sought to remove the incompatibilities
between quantum mechanics and relativity, by incorporating the approximate
nature of any physical measurement into quantum mechanics at a fundamental
level \rpi \rpii\ (and references therein). The resulting {\it
stochastic\/} quantum mechanics on phase space has significant advantages
over the usual formulation, for example, it gives a covariant, conserved
current whose time component is positive definite and so can be interpreted
as a position probability density. In extending this approach to quantum
mechanics and quantum field theory in curved spacetime, \prug\ has
introduced the Hilbert and Fock bundles over spacetime \rpii \rpnewp
\rpnewb, where the fibre above each point of spacetime is a copy of the
flat-space, stochastic Hilbert or Fock space, respectively.
In the first-quantised
theory, this avoids the problems of defining the Hilbert space and
operators appropriate to curved spacetime. It also means that the
probability density from the flat-space theory can be used to construct a
probability density in curved spacetime (see section 3 below). There is
however the new question of how a state in the fibre above any point is to
propagate to a fibre above another point, to which the present paper is
addressed. In geometro-stochastic quantum field theory, the use of the Fock
bundle implies that there is a unique vacuum state in the fibre above each
point, and a unique definition of a particle; the ambiguities of the usual
approach are thereby eliminated.

The bundle approach can also be applied to the conventional formulations of
quantum mechanics and quantum field theory simply by taking the standard
fibre to be the conventional flat-space Hilbert or Fock space, respectively. In
doing
this one loses the advantages peculiar to \prug's stochastic quantum
mechanics, such as the existence of a position probability density, but the
other quoted advantages of using the bundle are preserved.

The recent work of Graudenz \rgraud\ is also based on the use of a Fock
bundle over spacetime. However it differs from the present work in its
physical interpretation of the states and its treatment of their
propagation.

In the present paper we consider specifically the first-quantized theory of
a scalar particle of mass $m$ propagating on an arbitrary but fixed
background spacetime. (The background gravitational field is neither
quantized, nor is it altered by any back-reaction from the particle.) Our
aim is to obtain information about the quantum propagation law on the
Hilbert bundle over spacetime guided by the flat space limit of the
theory. In this limit, we require the bundle description to coincide with
\prug's (special-) relativistic stochastic quantum mechanics, which for
convenience we will often refer to as the {\it non-bundle\/}
description. Thus we take the non-bundle description as our reference
point, and, although we summarise this theory, our aim is not to give a
critical assessment of it.

The contents of the paper are as follows. In section 2 we give a
self-contained summary of the relevant structures and properties of the
relativistic stochastic quantum mechanical description of a free, spinless
particle of mass $m$. We also introduce the Hilbert bundle structure used
in the attempt to generalise this theory to a curved background spacetime.
In section 3 we summarise \prug's prescription for the physical
interpretation of the bundle formulation. In section 4 we discuss quantum
transport on the Hilbert bundle, describing it in terms of maps between the
fibres above pairs of points in spacetime. We define and discuss the
quantum mechanical propagator obtained from these maps, thus making contact
with the approach of \rpii\ and \rpnewb. We consider the flat space limit,
and by comparing with the non-bundle description we deduce the required
quantum transport law and propagator in this limit. In section 5 we discuss
parallel transport on the Hilbert bundle, and show that in the flat space
limit this coincides with quantum transport. This result is implicit in
certain statements of reference \rpnewb\ (e.g., the discussion following
eq.(4.6.8)), but does not appear to have been explicitly stated or proven.
In section 6 we obtain an integral representation for the quantum transport
propagator in the flat space limit. We compare this with the flat-space
limit of the quantum geometric propagator in curved spacetime defined in
eq.(4.6.7a,b) of \rpnewb, verifying that they agree. In section 7 we make
some remarks concerning the interpretation of the measure appearing in
\prug's work, verifying in detail that the quantum propagator in curved
spacetime is defined in a Poincar\'e gauge covariant manner. In section 8
we indicate the modifications to the foregoing needed to apply the bundle
formalism to the conventional description of a Klein-Gordon
particle. Section 9 contains the summary and conclusions.

Our notation is broadly compatible with that of references \rpii\ and
\rpnewb. However we do not follow the convention of using boldface and
normal fonts to distinguish certain objects represented by the same letter,
since with some fonts this can be unclear. Rather we use an underline to
make such distinctions.

\newsec{Relativistic stochastic quantum mechanics on phase space and the
Hilbert bundle over spacetime}

We consider a spinless particle of mass $m$ propagating in Minkowski
space. Let $k=(k^\mm)\in\R^4$, $\mm=0,1,2,3$, be the vector of components
of the particle's 4-momentum in an inertial frame. In terms of these
components the forward mass hyperboloid $\V_m^+$ is given by
\eqn\evmp{\V_m^+ = \{k\in\R^4\,|\,k^2 = m^2, k^0>0\}\ ,}
where $k^2 = k.k = k_\mm k^\mm = (k^0)^2 - ((k^1)^2+(k^2)^2+(k^3)^2)$.
We let $\V^+$ denote $\V_m^+$ in the case $m=1$.

The quantum description of this particle is based on the Hilbert space
$L^2(\V_m^+)$, consisting of complex functions $\tilde\psi$ on $\V_m^+$
which are square integrable with respect to the usual invariant measure
%$d\Omega_m$. In terms of the momentum components $k$ in an
%orthonormal frame this is
%\eqn\evmpmeas{d\Omega_m(k) = \delta(k^2-m^2)d^4k\ .}
\eqn\evmpmeas{d\Omega_m(k) = d^3k/(2k^0)\ .}
We let $d\Omega(k)$ denote $d\Omega_m(k)$ in the case $m=1$.
%The inner product is the one given by this measure, i.e.,
%\eqn\evmpip{\bra\tilde\psi_1|\tilde\psi_2\ket =
%\int_{\V_m^+}\tilde\psi_1^*(k)\tilde\psi_2(k)d\Omega_m(k)\ .}
%This space carries a representation of the Poincar\'e group by
%\eqn\evmppg{((b,\Lambda)\tilde\psi)(k) = e^{ib.k}
%\tilde\psi(\Lambda^{-1}k)\ .}

The phase space of the particle is a subspace of the cotangent bundle over
Minkowski space, which can be identified with a subspace of the tangent
bundle, isomorphic to $\R^4\times\V_m^+$, i.e., the set of all pairs
$(q,p)$, where $q\in\R^4$ are inertial coordinates on Minkowski space, and
$p\in\V_m^+$ as above \re. We will work with the coordinates $(q,v)$, where
$v=p/m\in\V^+$ is the vector of components of the particle's
4-velocity. Following \prug, it will be convenient to represent the point
$(q,v)$ by the complex variable
\eqn\ezeta{\zeta=q-i\ell v\in\R^4\times i\V^+\subset\C^4\ ,}
where $\ell$ is a real, positive constant\foot{In \prug's formulation of
geometro-stochastic quantum mechanics, $\ell$ is a fundamental non-zero
length scale, introducing a certain fuzziness in position and momentum
resolution, which has a regularizing effect. \prug\ suggests equating
$\ell$ with the Planck length, although for consistency with current
particle physics data any length of the order of $10^{-16}$cm or smaller
would be acceptable.}.

\prug\ gives a formulation of the quantum theory of the particle in which
$L^2(\V_m^+)$ is mapped to a Hilbert space $\F$ of functions on phase space
\rpi\ p.90, \rpii\ pp.82,138, \rpnewp\ eq.(2.2), \rpnewb\foot{The different
forms of \eqnn\efdef\efdef\ given in the cited references are reconciled by
the values of the overall normalisation coefficients used. Note that we
follow \rpnewp\ here, in particular in the adoption of $e^{-\ell v.k}$ for
the function playing the role of the momentum space resolution generator
$\eta$ in \rpi\ eq.(2.4.8) and $f$ in \rpii\ eq.(5.1.2). This choice is
made for convenience; our analysis is independent of the particular form
adopted for the resolution generator.}. That is, $\F$ consists of all
functions $\psi$ on phase space of the form
%\eqn\efdef{
%\psi(\zeta) = \tilde Z_{\ell,m}^{-1/2}\int_{\V_m^+} e^{-i\zeta.k}
%\tilde\psi(k)d\Omega_m(k)\ ,}
$$\psi(\zeta) = \tilde Z_{\ell,m}^{-1/2}\int_{\V_m^+} e^{-i\zeta.k}
\tilde\psi(k)d\Omega_m(k)\ ,\eqno\efdef$$
where $\tilde Z_{\ell,m}$ is a real constant depending on $\ell$ and $m$,
and $\tilde\psi\in L^2(\V_m^+)$. These wavefunctions satisfy the
Klein-Gordon equation
\eqn\efkg{(\partial_\mu\partial^\mu +m^2)\psi(\zeta) = 0\ ,}
where $\partial_\mu = \partial/\partial q^\mu$.
The inner product on $\F$ is defined by
\eqn\einpf{\bra\psi_1|\psi_2\ket =
\int_{\Sigma^+} \psi_1^*(\zeta)\psi_2(\zeta)
d\Sigma(\zeta) \ .}
This integral is carried out over any surface of the form
\eqn\esig{\Sigma^+ = \sigma\times\V^+ \ ,}
where $\sigma$ is an arbitrary Cauchy surface in Minkowski space
\rpi\ p.113, \rpii\ pp.87,140.
%\rpnewb\ p.117
The measure may be written as
%given in inertial coordinates $q$ and orthonormal components $p$ by
\eqn\efmeas{d\Sigma(\zeta) = 2v^\mm d\sigma_\mm(q)d\Omega(v) \ ,}
where $d\sigma_\mm(q)$ are the components of the
%unit normal 1-form to the surface $\sigma$ times the induced
volume element on $\sigma$ \re\foot{The
normalisation constant $\tilde Z_{\ell,m}$ in \efdef\ is chosen so that the
inner products on $L^2(\V_m^+)$ and $\F$ agree, under the map given by
\efdef.}.

The Poincar\'e group, \pg, is the set of all $g=(b,\Lambda)$, where
$b\in\R^4$ and $\Lambda\in\lg$, with the product
\eqn\epgprod{(b,\Lambda)(b'\Lambda') = (b+\Lambda b',\Lambda\Lambda')\ .}
Thus, for example,
\eqn\epginv{g^{-1}=(-\Lambda^{-1}b,\Lambda^{-1})\ .}
The action of \pg\ on Minkowski space is given by
\eqn\epgq{(b,\Lambda)q = \Lambda q - b\ ,}
where $q\in\R^4$ and the minus sign in front of $b$ arises because of the
adopted frame transformation given in eq.(22)
%%%% N.B. EQUATION NUMBER HARD-CODED ON LINE ABOVE!!!!!!!!!!!!!!!!!!!!!!!
below. It will be convenient to use the same action of $g$ on
complex vectors such as $\zeta=q-i\ell v$, i.e.,
\eqn\epgzeta{(b,\Lambda)\zeta=\Lambda\zeta-b
=\Lambda q-b-i\ell\Lambda v\ .}

The Hilbert space $\F$ carries a unitary, irreducible, spin zero,
phase-space representation $U(g)=U(b,\Lambda)$ of the Poincar\'e group by
\rpi\ p.92
%\eqn\eprep{(U(b,\Lambda)\psi)(\zeta) = \psi(\Lambda^{-1}(\zeta-b))\ .}
\eqn\eprep{(U(g)\psi)(\zeta) = \psi(g^{-1}\zeta) \ .}

There exists in $\F$ a unique, rotationally invariant wavefunction $\eta$,
called the (phase space) {\it resolution generator\/} of the representation
\rpi\ p.106\foot{In the detailed analysis of this reference, $\eta$ is the
phase space resolution generator corresponding to the momentum space
resolution generator $\tilde\eta\in L^2(V_m^+)$ via
$\eta=\W_\eta\tilde\eta$, where $\W_\eta$ is the map defined by
\efdef. Note that our $\eta$ and $\tilde\eta$ correspond respectively to
$\bar\eta$ and $\eta$ in this reference.}, \rpii\ p.82. Let $\Lambda_v$ be
the Lorentz boost to the frame of an observer moving with 4-velocity
$v$. For all $\zeta=q-i\ell v\in\R^4\times i\V^+$, define
\eqn\ephiz{\phi_\zeta = U(-q,\Lambda_v)\eta\ .}
Then
\eqn\ephizp{\psi(\zeta) =
\bra\phi_\zeta|\psi\ket\qquad\forall\psi\in\F \ .}
The states $\phi_\zeta$ constitute an overcomplete basis for $\F$ (a
coherent state basis). They give the following resolution of the identity
operator $\one$ on $\F$:
\eqn\eid{\one = \int_{\Sigma^+}|\phi_\zeta\ket \,
d\Sigma(\zeta)\,\bra\phi_\zeta| \ .}
$K^{(\ell)}$ defined by
\eqn\ekldef{K^{(\ell)}(\zeta',\zeta) = \bra\phi_{\zeta'}|\phi_{\zeta}\ket}
is the propagator on $\F$, i.e.,
\eqn\epropf{\psi(\zeta') =
\int_{\Sigma^+} K^{(\ell)}(\zeta',\zeta)\psi(\zeta) d\Sigma(\zeta)\ .}

For later reference, we consider the action of $U(g)$ on $\phi_\zeta$,
where $g=(b,\Lambda)\in\pg$:
\eqn\epgphiza{U(b,\Lambda)\phi_\zeta = U(b,\Lambda)U(-q,\Lambda_v)\eta =
U(b-\Lambda q,\Lambda\Lambda_v)\eta\ .} Using the facts that there exists
a rotation $\Lambda_R$ (Wigner rotation) such that $\Lambda\Lambda_v =
\Lambda_{\Lambda v}\Lambda_R$ (recall that $\Lambda_{\Lambda v}$ means the
boost of 4-velocity $\Lambda v$), and that $\eta$ is rotationally
invariant, this gives
\eqn\egphiz{U(g)\phi_\zeta = U(-(\Lambda q-b), \Lambda_{\Lambda v})\eta =
\phi_{\Lambda q-b-i\ell\Lambda v} = \phi_{g\zeta}\ ,}
where \ephiz\ and \epgzeta\ were used.

\prug\ gives a physical interpretation of a state $\psi\in\F$ as the
probability amplitude for simultaneous measurements of the {\it
stochastic\/} position and {\it stochastic\/} momentum of the particle \rpi\
p.97, \rpii\ pp.84,151,156. The amplitude squared $|\psi(\zeta)|^2$ of
$\psi$
%on any surface $q^0=\hbox{const}$
is to be interpreted as the probability density for the {\it mean\/}
stochastic position $q$ and the {\it mean\/} stochastic momentum $mv$ of
the particle. More precisely, the integral of this density over any Borel
set in the space $\{q-i\ell v\,|\,q\in\sigma,v\in\V^+\}$, where $\sigma$ is
a Cauchy surface in Minkowski space, gives the probability that a
measurement of the stochastic position and momentum at the ``time'' defined
by $\sigma$ will give a result lying in this Borel set.

The integral over all momenta of $|\psi(\zeta)|^2$ gives a
positive-definite density on configuration space, which is the zeroth
component of a conserved current \rpi\ p.111, \rpii\ p.87. Thus
%the integral of
this density can indeed be interpreted as a conserved,
configuration space probability
density. This of course is quite different from the
usual treatment of the relativistic particle via the Klein-Gordon equation,
where there is no conserved, positive-definite density on configuration
space.
%We note however that there does not appear to be any conserved {\it
%current\/} on {\it phase space\/} corresponding to $|\psi(\zeta)|^2$, so
%its interpretation as a {\it phase space\/} probability density may be
%criticized.

In the present work, we are primarily concerned with comparing the above
description of the relativistic particle with the Hilbert bundle
description summarised below. For this purpose we will compare
$\psi(\zeta)$ with the corresponding quantity in the bundle
description, specialising, however, to the flat space case. Equality of
these objects is certainly sufficient to ensure the agreement of the
physical predictions made by the two formulations.

We now summarise the Hilbert bundle structure used in the programme to
generalise the above to the case of a particle moving on a curved spacetime
\rpii\ \S5.1, \rpnewb\ \S\S4.1-2. Let $(M,g^L)$ be a Lorentzian 4-manifold. A
Poincar\'e frame $\us$ for $T_xM$, $x\in M$, is of the form
$\us=(\ua,\ue_\mm)$, $\mm=0,1,2,3$, where $\ua\in T_xM$ and $(\ue_\mm)$ is
an orthonormal frame (a Lorentz frame) for $T_xM$. The Poincar\'e frame
bundle $\PM$ over $M$ is the set of all $(x,\us)$, where $x\in M$ and
$\us$ is a Poincar\'e frame for $T_xM$. $\PM$ is a principal fibre bundle,
having as its structure group the Poincar\'e group $G=\pg$. Let
$a=\in\R^4$ be the vector of components $a^\mm$; $\mm=0,1,2,3$, and let
$\ua$ be defined, following
\rdi\foot{The minus sign in \eqnn\eacpt\eacpt\ is due to the fact that
$a=(a^\mm)$ is a vector in $T_xM$ pointing {\it towards\/} the origin of the
frame $(\ue_\mm)$ since the components $a^\mm$ parametrize an active
translation of this frame as given by eq.(22) in the text.
%%%% N.B. EQUATION NUMBER HARD-CODED ON LINE ABOVE!!!!!!!!!!!!!!!!!!!!!!!
Eq.\eacpt\ then defines a vector $\ua$ pointing from the
origin of the Poincar\'e frame $s=(\ua,e_\mm)$ to the point of contact
between $T_xM$ and $M$. Thus it is opposite to the convention adopted in
\rpii\ and \rpnewb\ for the quantity denoted by boldface $a$.
For all other vectors $\uv$ referred to the frame $(\ue_\mm)$ we
write $\uv=v^\mm\ue_\mm$.}\footlab\facpt, by
%\eqn\eacpt{\ua=-a^\mm\ue_\mm\ .}
$$\ua=-a^\mm\ue_\mm\ .\eqno\eacpt$$
Then the right action $\ra_{g^{-1}}$ of
$g^{-1}=(-\Lambda^{-1}b,\Lambda^{-1})\in\pg$ on $\PM$ is
\eqn\era{\ra_{g^{-1}}(x,\ua,\ue_\mm) =
(x,\underline{\Lambda a+b}\,,\ue_\nn(\Lambda^{-1})^\nn{}_\mm)\ ,}
where $\underline{\Lambda a+b}=-(\Lambda a+b)^\mm\ue'_\mm$, with
$\ue'_\mm=\ue_\nn(\Lambda^{-1})^\nn{}_\mm$.
%The right action $\ra_g$ of $g=(b,\Lambda)\in\pg$ on $\PM$
%is\foot{Reference \rdi\ gives a more elegant expression of $\ra_g$, but
%note that \eqnn\era\era\ would there be considered to be the action of
%$g^{-1}$.}
%%\eqn\era{\ra_g(x,\ua,\ue_\mm) =
%%(x,\ua+b^\mm\ue_\mm,\ue_\nn\Lambda^\nn{}_\mm) \ .}
%$$\ra_g(x,\ua,\ue_\mm) =
%(x,\ua+b^\mm\ue_\mm,\ue_\nn\Lambda^\nn{}_\mm) \ .\eqno\era$$

Let the {\it Hilbert bundle\/} $\hb$ over $M$ be a fibre bundle associated to
$\PM$, having as its typical fibre the 1-particle Hilbert space $\F$.
$\hb$ may be viewed as the $G$-product $\PM\times_G\F$ of $\PM$ and $\F$,
i.e., the quotient of $\PM\times\F$ by an action of $G$, defined as
follows. Let $x\in M$, $\us$ be a Poincar\'e frame for $T_xM$, and
$\psi\in\F$, so $(x,\us)\in \PM$ and $(x,\us,\psi)\in \PM\times\F$. An
action of $\pg$ on $\PM\times\F$ is defined by
%\eqn\eraprod{g:(x,\us,\psi)\mapsto (R_{g}(x,\us), U(g^{-1})\psi)
%\qquad\forall g\in\pg\ .}
\eqn\eraprod{(x,\us,\psi)\mapsto (R_{g^{-1}}(x,\us), U(g)\psi)
\qquad\forall g\in\pg\ .}
Let $[(x,\us,\psi)]$ be the equivalence class of $(x,\us,\psi)$ under this
action. The equivalence classes $[(x,\us,\psi)]$ are the elements of
$\hb$. The projection $\pi$ on $\hb$ is the obvious one,
\eqn\eproj{\pi:\hb\rightarrow M
\qquad\hbox{by}\qquad\pi:[(x,s,\psi)]\mapsto x\ .}

For all $x\in M$ and all Poincar\'e frames $\us$ for $T_xM$, we can
define the maps
\eqn\ecnmap{\usig_x^{\us} : \pi^{-1}(x) \rightarrow\F \qquad
\hbox{by}\qquad\usig_x^{\us} : [(x,\us,\psi)] \mapsto \psi
\qquad\forall\psi\in\F\ .}
Note that each equivalence class in $\pi^{-1}(x)$ has exactly one element
$(x,\us',\psi)$ such that $\us'=\us$, so $\usig_x^{\us}$ is indeed defined
and single-valued on all of $\pi^{-1}(x)$. Let $\us$ and $\us'$ be two
Poincar\'e frames for $T_xM$ and $g\in\pg$ such that
%\eqn\egdef{(x,\us') = \ra_g(x,\us)\ .}
\eqn\egdef{(x,\us') = \ra_{g^{-1}}(x,\us)\ .}
It follows from \eraprod\ and \ecnmap\ that
%\eqn\esigt{\usig_x^{\us'} = U(g^{-1})\,.\,\usig^{\us}_x\ .}
\eqn\esigt{\usig_x^{\us'} = U(g)\,.\,\usig^{\us}_x\ .}

A Poincar\'e frame $\us$ for $T_xM$ associates to each
$\upsi_x\in\pi^{-1}(x)$ an element $\Psi_x^{\us}$ of the standard fibre $\F$
by
\eqn\ewvfun{\Psi_x^{\us} = \usig_x^{\us}\upsi_x \ .}
$\Psi_x^{\us}$ is called the {\it wavefunction\/} of the state $\upsi_x$ in
the {\it Poincar\'e gauge\/} $\us$. This induces an inner product on the
fibres of the bundle $\hb$ by
\eqn\einpre{\bra\upsi'_x|\upsi_x\ket =
\bra\Psi_x^{\prime\us}|\Psi_x^\us\ket
%= \bra\usig_x^\us\upsi'_x|\usig_x^\us\upsi_x\ket
\qquad\forall\,\upsi'_x,\upsi_x\in\pi^{-1}(x)\ ,}
where $\us$ is any Poincar\'e frame for $T_xM$, and the inner product on
the right hand side is the one on the standard fibre defined by equation
\einpf. It follows from \esigt\ that $\bra\upsi'_x|\upsi_x\ket$ is
independent of the choice of the frame $\us$ appearing in \einpre.

For all $x\in M$, all Poincar\'e frames $\us$ for $T_xM$, and all $\zeta =
q-i\ell v\in\R^4\times i\V^+$ define
\eqn\euphiz{\uphi^\us_\zeta = (\usig_x^\us)^{-1}\phi_\zeta
\qquad(\in\pi^{-1}(x))\ ,}
where $\phi_\zeta$ is a basis state in $\F$ defined by \ephiz. Combining
\ephizp, \ewvfun\ and \euphiz, it follows that the value of the
wavefunction $\Psi_x^\us$ (corresponding to the state $\upsi_x$ in the
Poincar\'e gauge $\us$) at $\zeta$ is given by
\eqn\euphizp{\Psi_x^\us(\zeta) = \bra\uphi_\zeta^\us|\upsi_x\ket
\qquad\forall\,\upsi_x\in\pi^{-1}(x)\ .}
{}From the properties of the coherent states $\phi_\zeta$, it follows that
the $\uphi^\us_\zeta$ for any $\us$ constitute an overcomplete basis for
$\pi^{-1}(x)$, and from \eid\ they give the following resolution of the
identity map $\onex$ on $\pi^{-1}(x)$:
\eqn\exid{\onex = \int_{\Sigma^+}|\uphi_\zeta^\us\ket \,
d\Sigma(\zeta)\,\bra\uphi_\zeta^\us| \ .}
Let $x$, $\us$, $\us'$ and $g$ be as in \egdef. The gauge-dependence of the
$\uphi^\us_\zeta$ is shown by
%\eqn\euphig{\uphi^{\us'}_\zeta = (\usig_x^{\us'})^{-1}
%\phi_\zeta = (\usig_x^{\us})^{-1}.U(g)\phi_\zeta = (\usig_x^{\us})^{-1}
%\phi_{g\zeta} = \uphi^{\us}_{g\zeta}\ ,}
\eqn\euphig{\uphi^{\us'}_\zeta = (\usig_x^{\us'})^{-1}
\phi_\zeta = (\usig_x^{\us})^{-1}.U(g^{-1})\phi_\zeta = (\usig_x^{\us})^{-1}
\phi_{g^{-1}\zeta} = \uphi^{\us}_{g^{-1}\zeta}\ ,}
where \esigt\ and \egphiz\ were used.

\newsec{Physical interpretation of sections of the Hilbert bundle}

In the Hilbert bundle description, the physical state of the 1-particle
system is described by a section $\upsi$ of the bundle $\hb$, i.e.,
\eqn\esec{\upsi:M\rightarrow \hb\ ,
\qquad\upsi:x\mapsto\upsi_x\,\in\,\pi^{-1}(x)\ .}
Thus this description involves a vector $\upsi_x$ in the fibre
$\pi^{-1}(x)$ above each point $x\in M$, whereas in the non-bundle
description the state is described by a single vector $\psi\in\F$. This
apparent oversupply of information contained in such a section is removed
in its physical interpretation, because each state $\upsi_x$ determines the
stochastic
%position and momentum
phase space
probability amplitude only at the corresponding base point
$x=\pi(\upsi_x)$.
%(More generally, it determines this density
%approximately in a region around $x$.)

The phase space of the particle is (up to a factor of $m$ in the momentum)
the set of all pairs $(x,\uv)$, where $x\in M$ and $\uv\in\V_x^+$, where
$\V_x^+$ is the forward 4-velocity hyperboloid at $x$, i.e.,
\eqn\evxp{\V_x^+ = \{\uv\in T_xM\ |\ g^L(\uv,\uv)=1,
\,\uv\ \hbox{future directed}\}\ .}
The phase space probability amplitude $\Psi(x,\uv)$ predicted by the
section $\upsi$ is defined as follows \rpii\ p.156. Let $\us$ be a
section\foot{For convenience, we make the usual assumption that $\PM$
admits a global section. This is physically desirable since it is
associated with the existence of a spin structure on spacetime.} of $\PM$,
i.e.,
\eqn\esecpm{\us:M\rightarrow\PM\qquad\hbox{by}
\qquad x\mapsto\us(x) = \left(\ua(x), \ue_\mm(x)\right)\ .}
Thus $\us$ provides a Poincar\'e gauge choice $\us(x)$ at each $x\in M$,
and we shall also refer to the section $\us$ as a choice of gauge. Let
$a(x)=(a^\mm(x))\in\R^4$ be the vector of components $a^\mm(x)$ associated
to $\ua(x)$ as defined in \eacpt,
%\eqn\eacmp{\ua(x) = a^\mm(x)\ue_\mm(x) \ .}
%$$\ua(x) = a^\mm(x)\ue_\mm(x) \ .\eqno\eacmp$$
and let $v=(v^\mm)\in\V^+$ be the vector of components of $\uv$ defined by
$\uv = v^\mm\ue_\mm(x)$ (see footnote \facpt). Then the phase space
probability amplitude predicted by $\upsi$ is
\eqn\eprob{\Psi(x,\uv) = \Psi_x^{\us(x)}(\hat\zeta(x)) = (\usig_x^{\us(x)}
\upsi_x)(\hat\zeta(x))\ ,}
where
%\eqn\ezethat{\hat\zeta(x) = -a(x)-i\ell v\ \in\ \R^4\times i\V^+\ .}
\eqn\ezethat{\hat\zeta(x) = -a(x)-i\ell v\ \in\ \R^4\times i\V^+\ ,}
with $q^\mm=-a^i(x)$ denoting the point of contact in $T_xM$.
%Note that the sign of $a(x)$ in this equation is opposite to that in
%\rpii~eq.(5.5.1a), because of the sign difference in our equation \eacpt\
%compared with \rpii~eq.(2.3.11).
It follows from \era, \esigt\ and
\ewvfun\ that $\Psi(x,\uv)$ is independent of the choice of the gauge $\us$
appearing in \eprob\foot{There is an interpretation, which we will not
use, of $\upsi_x$ as a (gauge-independent) function on the space
$T_xM\times\V_x^+$, i.e., $\upsi(\uq,\uv)\in\C$, where $\uq\in T_xM$ and
$\uv\in\V_x^+$ \rpnewb\
\S4.2. In terms of this interpretation, \eprob\ takes the form
$\Psi(x,\uv)=\upsi_x(\ua(x),\uv)$, i.e., $\Psi(x,\uv)$ is obtained by
restricting $\upsi_x$ to the point of contact $\uq=\ua(x)$ between $T_xM$
and $M$.}.\footlab\fftxmvx\ Analogously to the non-bundle description, the
amplitude squared $|\Psi(x,\uv)|^2$ of $\Psi(x,\uv)$ would be interpreted
as the probability density on phase space.

A result of \prug\ (discussion in \rpnewp\ below eq.(3.5) and \rpnewb\
p.178) states in the current context that if $\Psi_x^{\us(x)}(-a(x)-i\ell
v)$ for all $v\in\V^+$ is fixed, then $\Psi_x^{\us(x)}(q-i\ell v)$ for all
$q\in\R^4$ is determined by analyticity. Combining this with \eprob, we see
that there is actually a one-to-one correspondence between sections of $\hb$
and phase space probability amplitudes, i.e., not only does $\upsi_x$
determine $\Psi(x,\uv)$ via \eprob, but also $\Psi(x,\uv)$ (for all $\uv$)
determines $\upsi_x$ by the result just quoted.

\prug\ also gives a stronger probability interpretation of a section
$\upsi$, \rpnewb\ eq.(4.2.16), where $\upsi_x$ determines the relative
probability density $\Psi(x',\uv)$ at points $x'$ near to $x$, up to some
accuracy related to the curvature of $(M,g^L)$\foot{This is based on the
interpretation of $\upsi_x$ as a function on $T_xM\times\V_x^+$ mentioned
in footnote \fftxmvx. It uses the exponential map at $x$ to associate $x'$
with some $\uq\in T_xM$, and then determines $\Psi(x',\uv)$ from the
restriction of $\upsi_x$ to this value of $\uq$.}\footlab\ffxp.
This interpretation is
stronger because it requires an approximate consistency between $\upsi_x$
and $\upsi_{x'}$ at nearby points $x$ and $x'$. We will not base what
follows on this interpretation, although we will make a comment on it at
the end of the next section.

\newsec{Quantum transport in the Hilbert bundle}

Sections 5.4 of \rpii\ and 4.6 of \rpnewb\ postulate explicit path
integral-like formulae for ``geometro-stochastic'' and
``quantum-geometric'' forms of the quantum propagator in the bundle over
curved spacetime. In this section we give a rather general definition of
quantum transport and the associated propagator, without however
postulating any explicit form in curved spacetime, and then go on to
consider their specific forms in the flat space case.

In the non-bundle description, the propagation of the wavefunction in time
is governed by the Klein-Gordon equation. This is supposed to be paralleled
in the Hilbert bundle formalism by a law for the propagation of a state
from the fibre above one point in the base to the fibre above another
point, called {\it quantum transport\/}. That is, for pairs of points
$x',x\in M$, there is a map
\eqn\eqdef{Q(x',x):\pi^{-1}(x)\rightarrow\pi^{-1}(x')\ ,
\qquad Q(x',x):\upsi_x\mapsto Q(x',x)\upsi_x
%\,\in\,\pi^{-1}(x')
\ ,}
which implements quantum mechanical propagation in the bundle description.
Thus, given an ``initial'' state $\upsi_x$ in the fibre above some point
$x$, the maps $Q(x',x)$ can be used to propagate this state to other points
$x'$, building up a section $\upsi$ of $\hb$ (refer to equation \esec) by
\eqn\esecqdef{\upsi_{x'}=Q(x',x)\,\upsi_x\ .}
No path dependence in $Q(x',x)$ for a path joining $x'$ and $x$ will be
assumed, nor will any particular single path be considered.
A priori we might expect that the $Q$ maps will not be defined for all
pairs $(x',x)$, but that for causality reasons they might only be defined
when $x'$ is in the causal future of $x$. In this case the resulting
section $\upsi$ would only be defined on part of $M$. We will return to
this point below.

The {\it quantum transport propagator\/} $\uk^\us(x',\zeta';x,\zeta)$ from
$x$ to $x'$ in the arbitrary gauge $\us$ is defined by\foot{The definition
of $\uk^\us(x',\zeta';x,\zeta)$ in terms of $Q(x',x)$ is given here in
analogy to the definition of the parallel transport propagator in terms of
the parallel transport map in \rpii\ \S 5.4, which, of course, is path
dependent in curved spacetime.}
\eqn\eqprop{\uk^\us(x',\zeta';x,\zeta)=
\bra\uphi_{\zeta'}^{\us(x')}|\,Q(x',x)\,|\uphi_{\zeta}^{\us(x)}\ket\ ,}
i.e., this is the matrix element of the map $Q(x',x)$ between the fibre
basis states $\uphi_{\zeta'}^{\us(x')}\in\pi^{-1}(x')$ and
$\uphi_{\zeta}^{\us(x)}\in\pi^{-1}(x)$ defined in \euphiz. Taking the inner
product of both sides of \esecqdef\ with $\uphi_{\zeta'}^{\us(x')}$, and
using \euphizp, \exid\ and \eqprop\ we find
\eqn\eqwfprop{\Psi_{x'}^{\us(x')}(\zeta') =
\int_{\Sigma^+}\uk^\us(x',\zeta';x,\zeta)\,\Psi_x^{\us(x)}(\zeta)\,d
\Sigma(\zeta)\ ,}
i.e., $\uk^\us(x',\zeta';x,\zeta)$ is the quantum propagator from $x$ to
$x'$ for the wavefunction of the state in the gauge $\us$. (Note that since
this expression is written in terms of wavefunctions, i.e., elements of the
standard fibre $\F$, the measure $d\Sigma(\zeta)$ is that defined by
\einpf, and $\Sigma^+$ is the surface defined by \esig.)
Let $\us$ and $\us'$ be two gauges, and define $g(x)\in\pg$ by
%\eqn\egxdef{(x,\us'(x)) = \ra_{g(x)}(x,\us(x))\ .}
\eqn\egxdef{(x,\us'(x)) = \ra_{g(x)^{-1}}(x,\us(x))\ .}
The gauge-dependence of $\uk^\us$ is shown by
%\eqn\eqprg{\eqalign{\uk^{\us'}(x',\zeta';x,\zeta)=
%\bra&\uphi_{\zeta'}^{\us'(x')}|\,Q(x',x)\,|\uphi_{\zeta}^{\us'(x)}\ket
%\cr=\bra&\uphi_{g(x')\zeta'}^{\us(x')}|\,Q(x',x)\,|
%\uphi_{g(x)\zeta}^{\us(x)}\ket=
%\uk^{\us}(x',g(x')\zeta';x,g(x)\zeta)\ ,\cr}}
\eqn\eqprg{\eqalign{\uk^{\us'}(&x',\zeta';x,\zeta)=
\bra\uphi_{\zeta'}^{\us'(x')}|\,Q(x',x)\,|\uphi_{\zeta}^{\us'(x)}\ket
\cr=\bra&\uphi_{g(x')^{-1}\zeta'}^{\us(x')}|\,Q(x',x)\,|
\uphi_{g(x)^{-1}\zeta}^{\us(x)}\ket=
\uk^{\us}(x',g(x')^{-1}\zeta';x,g(x)^{-1}\zeta)\ ,\cr}}
where \euphig\ and \eqprop\ were used. Thus the quantum propagator is
dependent on the gauge $\us$ only at the points $x'$ and $x$.

For the rest of this section we will concentrate on the special case where
the spacetime base $(M,g^L)$ is Minkowski space. In this case, the physical
interpretation of a section of $\hb$ given by \eprob\ and \ezethat\ implies a
restriction on the sections of $\hb$ which are allowed. This is because in
flat space, the Hilbert bundle description must reproduce the set of
possible phase space probability densities given by the non-bundle
description. That is, the only sections $\upsi$ of $\hb$ allowed are those
which predict an amplitude $\Psi(x,\uv)$ which corresponds to one of the
states $\psi\in\F$. This in turn gives information on the quantum transport
law in the case when the base spacetime is Minkowski space, because the
sections of $\hb$ which it generates via \esecqdef\ must satisfy this
restriction. In the remainder of this section we give the corresponding
quantum transport law and discuss some of its properties.

Since the base manifold $(M,g^L)$ is Minkowski space, we can introduce
inertial coordinates. {}From now on we will use $x=(x^\mm)\in\R^4$ to denote
the coordinates of a point in $M$ in some inertial coordinate system. At
each point $x\in M$ we define an orthonormal frame $(\ue^L_\mm(x))$ by
\eqn\elorfrm{\ue^L_\mm(x) = {\partial\phantom{x^\mm}\over\partial x^\mm}
\bigg|_x \ ,}
This of course is just the natural frame in our chosen coordinates $x$, and
$\ue^L_\mm(x)$ is of course equal to the parallel transport of
$\ue^L_\mm(O)$ from some fixed point $O\in M$ to $x$. Define the section
$\us_L$ of $\PM$ by
\eqn\elorsec{\us_L(x) = (\,\ua_L(x)=x^\mm\ue^L_\mm(x),\,\ue^L_\mm(x)\,)
\qquad\forall x\in M\ .}

Let us insert some general remarks at this place. One can obtain any frame
$s(x)=(\ua(x),e_\mm(x))$ in $T_xM$ by a Poincar\'e transformation applied
to the frame\foot{One can consider $x\in\R^4$ to be the vector of
components $x^\mm$ of a point $x\in M$ in a basis $(\ue_\mm(O))$ for $M$,
considered as a vector space with origin $O$. The frame $(\ue_\mm(O))$ for
$M$ is naturally identified with the Poincar\'e frame
$\so=(0,\ue^L_\mm(O))$, and hence
with the frames $\us_L(x)=(x^\mm\ue^L_\mm(x),\ue^L_\mm(x))$ for all $x$,
which all have their origins displaced to the point $O$ \rpii~p.43,
\rpnewb~eq.(4.1.13). (Compare also footnote \facpt\ in this context.)}
%The Poincar\'e frames $s_L(x)=(-x^\mm\ue^L_\mm(x),\ue^L_\mm(x))$ for all
%$x$ can then naturally be identified with the $(\ue_\mm(O))$ for $M$
%via the exponential map at $\ux=x^\mm\ue_\mm(O)$.}
\footlab\fxbar
$\so=(0,\ue^L_\mm(O))$ in $T_OM$ according to
\eqn\exxx{(x,s(x))=\ra_{\tilde g_{s(x)}^{-1}}(O,\so)}
where $\tilde g_{s(x)}^{-1}=(x+a(x),\Lambda(x))$. Eq.\exxx\
is similar to
\era\ or \egxdef, however, when the spacetime base is Minkowskian the
Poincar\'e frames associated with different base points (here $O$ and $x$)
can also be related by Poincar\'e transformations (denoted by a tilde) with
translations given by $\tilde a(x)=x+a(x)$.

Let us write $\tilde g_{s(x)}$ as a product $\tilde
g_{s(x)}=g_{\Lambda(x)}\tilde g_{(x+a(x))}$ composed of a pure translation
given by $(x+a(x),1)$ and a pure Lorentz transformation given by
$(0,\Lambda(x))$. We can then construct a frame in all the tangent spaces
$T_xM$ which is parallel to the fixed Lorentz frame $e^L(O)$ at $O\in M$
and define a corresponding ``parallel'' Poincar\'e gauge by
\eqn\eyyy{(x,s^{||}(x))=\ra_{\tilde g_{(x+a(x))}}(O,\so)\ .}
For $a(x)=a_L(x)=-x$
this yields the global Lorentz gauge \elorsec\ on $\PM$. However, the most
general Poincar\'e frame $s(x)=(\ua(x),e_\mm(x))$ in $T_xM$ and the
corresponding general gauge on $\PM$ is obtained from the parallel gauge by
an arbitrary $x$-dependent smooth $\Lambda$-transformation in $x\in M$
\eqn\ezzz{(x,s(x))=\ra_{g_{\Lambda(x)}^{-1}}(x,s^{||}(x))}
generating the most general section of $\PM$ in the Minkowski case.

A physically acceptable quantum propagation law is now defined by
\eqn\eqt{Q(x',x) = (\usig_{x'}^{\us_L(x')})^{-1}.\usig_x^{\us_L(x)}\ .}
The proof that these $Q$ maps do generate an acceptable section $\upsi$ of
$\hb$ from any given ``initial'' state $\upsi_x\in\pi^{-1}(x)$ is as follows.
The value $\upsi_{x'}$ of $\upsi$ at $x'$ is
\eqn\eqsec{\upsi_{x'} = Q(x',x)\upsi_x = (\usig_{x'}^{\us_L(x')})^{-1}.
\usig_x^{\us_L(x)}\upsi_x = (\usig_{x'}^{\us_L(x')})^{-1}
\Psi_x^{\us_L(x)}\ ,}
where \ewvfun\ was used. Inserting the choice $\us=\us_L$ in \eprob, we
find that the phase space probability amplitude predicted by this section
is
\eqn\eqsecpz{\Psi(x',\uv) =
(\usig_{x'}^{\us_L(x')}\upsi_{x'})(\hat\zeta(x')) =
\Psi_x^{\us_L(x)}(\hat\zeta(x'))\ ,}
where \eqsec\ was used, and where in the gauge $\us=\us_L$
\ezethat\ gives
\eqn\eqzethat{\hat\zeta(x') = -a_L(x')-i\ell v = x'-i\ell v\ .}
Hence we find that
\eqn\eqsecp{\Psi(x',\uv) = \Psi_x^{\us_L(x)}(x'-i\ell v)\ ,}
i.e., the phase space probability amplitude predicted by $\upsi$
corresponds to the state $\Psi_x^{\us_L(x)}\in\F$ in the non-bundle
description. Thus $\upsi$ does indeed satisfy the requirement that it
should correspond to a state in the non-bundle description, as discussed in
the paragraph following equation \eqprg. Further, because of the
observation at the end of the previous section that there is a one-to-one
correspondence between phase space probability amplitudes and sections of
$\hb$, the quantum transport law defined by \eqt\ is the {\it unique\/} law
which satisfies this requirement of compatibility with the non-bundle
description.

We note that this flat-space quantum transport law satisfies the
composition law
\eqn\eqtcmp{Q(x',x) = Q(x',x_1).Q(x_1,x)\ ,}
for any $x_1\in M$.
%(such that $Q(x',x_1)$ and $Q(x_1,x)$ are defined).
Substituting this and \exid\ in \eqprop\ gives
\eqn\eqpcmp{\uk^\us(x',\zeta';x,\zeta)=
\int_{\Sigma^+}\uk^\us(x',\zeta';x_1,\zeta_1)\,\uk^\us(x_1,\zeta_1;x,\zeta)
\,d\Sigma(\zeta_1)\ ,}
for any $x_1\in M$.

We now derive an expression for the quantum transport propagator defined by
\eqprop, for the quantum transport law defined by \eqt. For the flat space
case, it would be sufficient and indeed simplest to use only the gauge
$\us_L$. However, with a view to our later discussion of the generalisation
to curved space, we consider a fully general gauge $\us$, which may differ
from $\us_L$ by a general Poincar\'e transformation, i.e., both by a
Lorentz transformation of the frame $(\ue^L_\mm(x))$ and by a shift of its
origin in $T_xM$. Hence, let $\us$ be an arbitrary gauge as in \exxx\
or \ezzz, and let
$g_\us(x)=(b(x),\Lambda(x))\in\pg$, then one obtains a general frame
$s(x)=(\underline a(x),e_\mm(x))$, with $a(x)=\Lambda(x)a_L(x)+b(x)$ and
$e_\mm(x)=e^L_\nn(x)[\Lambda(x)^{-1}]^\nn{}_\mm$, from $s_L(x)$ in the
following manner
%be defined by
%\eqn\egs{(x,\us(x))=\ra_{g_\us(x)}(x,\us_L(x))\ .}
\eqn\egs{(x,\us(x))=\ra_{g_\us(x)^{-1}}(x,\us_L(x))\ .}
Then
%\eqn\eqpqt{\eqalign{\uk^\us(x',\zeta';x,\zeta)=
%\bra&\phi_{\zeta'}|\,\usig_{x'}^{\us(x')}.(\usig_{x'}^{\us_L(x')})^{-1}.
%\usig_x^{\us_L(x)}.(\usig_x^{\us(x)})^{-1}\,|\phi_\zeta\ket\cr=
%\bra&\phi_{\zeta'}|\,U(g_\us(x')^{-1})U(g_\us(x))\,|\phi_\zeta\ket\ ,\cr}}
\eqn\eqpqt{\eqalign{\uk^\us(x',\zeta';x,\zeta)=
\bra&\phi_{\zeta'}|\,\usig_{x'}^{\us(x')}.(\usig_{x'}^{\us_L(x')})^{-1}.
\usig_x^{\us_L(x)}.(\usig_x^{\us(x)})^{-1}\,|\phi_\zeta\ket\cr=
\bra&\phi_{\zeta'}|\,U(g_\us(x'))U(g_\us(x)^{-1})\,|\phi_\zeta\ket\ ,\cr}}
where \euphiz\ and \esigt\ were used. For example, in the gauge $\us=\us_L$
defined by \elorsec, we have of course $g_{\us}(x)=1\ \forall x$, so that
\eqn\eqpqtsl{\uk^{\us_L}(x',\zeta';x,\zeta)=
\bra\phi_{\zeta'}|\phi_\zeta\ket = K^{(\ell)}(\zeta',\zeta)\ ,}
where $K^{(\ell)}(\zeta',\zeta)$ is the propagator in the non-bundle
description defined in \ekldef. Thus, in particular, we note that
$\uk^{\us_L}(x',\zeta';x,\zeta)$ is actually independent of $x'$ and $x$.

We close the section with some comments. As mentioned in the previous
section, \prug\ also gives a stronger physical interpretation of a section
of $\hb$ (\rpnewb\ eq.(4.2.16)) than the one we have relied on here. In the
flat space case, this interpretation allows the probability density
$\Psi(x',\uv)$ for all $x'\in M$ to be calculated exactly from $\upsi_x$,
and so it implies an exact consistency requirement between $\upsi_x$ and
$\upsi_{x'}$ for all $x,x'\in M$. It can be shown that the sections $\upsi$
generated by the quantum transport law defined by $\eqt$ do satisfy this
consistency condition.

At the beginning of this section we raised the point that the maps
$Q(x',x)$ may not necessarily be defined for all pairs of points $x',x\in
M$. However it is clear that the maps defined by $\eqt$ are well-defined
for all such pairs of points, and that they satisfy \eqtcmp\ without
specifying a path. This is physically reasonable in the flat-space case,
since the fibre above each point $x$ is an exact copy of the non-bundle
description of the particle. Thus the state $\upsi_x$ for any $x$ actually
contains a complete description of the system, and it is therefore not
surprising that it determines the states $\upsi_{x'}$ at all other points
$x'$, not only at points in the causal future of $x$. This also raises the
question of the physical status of the quantum transport law on the bundle.
Should it be thought of as being analogous to the Klein-Gordon equation in
that it determines some sort of causal, dynamical propagation of the state
on the bundle, or is it more of the nature of a kinematical consistency
condition, resulting from the geometrical structure of the Hilbert bundle
formalism?

\newsec{Parallel transport in the Hilbert bundle}

We begin by summarising the relevant parts of the treatment of parallel
transport given in \rpii\ \S\S5.2--4 and \rpnewb\ \S\S4.3--4,4.6, before
going on to consider the flat space case.

The Levi-Civita connection $\uomega'$ on the Lorentz frame bundle over $M$,
consistent with the metric $g^L$, can be uniquely extended to an affine
connection $\tilde\uomega$ on the Poincar\'e frame bundle $\PM$ over $M$
\rk, \rkn~\S III.3. The result of this can be summarised as follows. Let
$\uu$ be a section of the Lorentz frame bundle, i.e.,
$\uu:x\mapsto\uu(x)=(\ue_\mm(x))\ \forall x\in M$, where $(\ue_\mm(x))$ is
an orthonormal frame for $T_xM$. Then the usual connection 1-forms in the
{\it Lorentz gauge\/} $\uu$ are given by the pull-back $\uu^*\uomega'$ of
$\uomega'$ to $M$ under $\uu$. Let $\bar\us$ be the section of $\PM$
defined by
\eqn\elsecpm{\bar\us(x) = (0, \uu(x))\ .}
Then the pull-back of $\tilde\uomega$ to $M$ under this section of $\PM$ is
\eqn\etilompb{\bar\us^*\tilde\uomega = (\utheta^\mm,
\uu^*\uomega')\ ,}
where $(\utheta^\mm)$ is the coframe dual to the frame $(\ue_\mm)$. The
pull-back $\us^*\tilde\uomega$ of $\tilde\uomega$ under a general
Poincar\'e gauge $\us$ (equivalently, the value of $\tilde\uomega$ on all
of $\PM$) follows by the usual gauge transformation properties, and yields
$\us^*\tilde\omega=(\tilde\utheta^\mm,\uu^*\uomega')$ with
$\tilde\utheta^\mm = \utheta^\mm+\nabla a^\mm$ (compare \rdi). Different
extensions of $\uomega'$ to a connection on $\PM$ are thus obtained by
replacing the $\utheta^\mm$ in \etilompb\ by some other 1-forms
$\tilde\utheta^\mm$ (soldering forms).

Prugovecki chooses to use $\tilde\uomega$ as the connection on $\PM$
corresponding to the metric $g^L$ on $M$. Thus the additional components of
$\tilde\uomega$ compared to $\uomega'$ are regarded as a mathematical
device, rather than as being physical fields which might be influenced by
matter via some source equation. They are chosen to be certain fixed
1-forms on $\PM$, which merely provide a means of extending the Levi-Civita
connection to $\PM$ and so defining parallel transport on $\hb$. So,
essentially, $\bar\us^*\tilde\uomega$ defined in \etilompb\ is used. The
association of $\hb$ to $\PM$ then determines a parallel transport law on $\hb$
in the usual way (e.g., \ryellow\ p.369). Thus for any pair of points
$x',x\in M$ and any curve $\gamma$ joining them, there is a corresponding
{\it parallel transport map}
\eqn\epltr{\tau_\gamma(x',x):\pi^{-1}(x)\rightarrow\pi^{-1}(x')
\ ,\qquad\tau_\gamma(x',x):\upsi_x\mapsto\tau_\gamma(x',x)\,\upsi_x\ .}
Let $\us$ be some Poincar\'e frame for $T_xM$, and
$\uphi_\zeta^\us\in\pi^{-1}(x)$ be one of the corresponding basis states
defined by \euphiz. Then from \ecnmap\ and the definition of parallel
transport in an associated bundle, it follows that
\eqn\eptuphiz{\tau_\gamma(x',x)\,\uphi_\zeta^\us =
\uphi_\zeta^{\tau_\gamma(x',x)\us}\ ,}
where on the right hand side $\tau_\gamma(x',x)$ is now used to denote
parallel transport in the principal bundle $\PM$ under the connection
$\tilde\uomega$, i.e., $\tau_\gamma(x',x)\us$ denotes the result of the
parallel transport of $\us$ from $x$ to $x'$ along $\gamma$. By the use of
\exid, an arbitrary state $\upsi_x\in\pi^{-1}(x)$ can be expanded in terms
of the $\uphi_\zeta^\us$ for any $\us$. Therefore the parallel transport of
an arbitrary state can be written as
\eqn\eptupsix{\tau_\gamma(x',x)\,\upsi_x =
\int_{\Sigma^+}\Psi_x^\us(\zeta)\,\uphi_\zeta^{\tau_\gamma(x',x)\us}d
\Sigma(\zeta)\ ,}
where $\us$ is any Poincar\'e frame for $T_xM$ and $\Psi_x^\us(\zeta)$ is
given by \euphizp.

The {\it parallel transport propagator\/} $K_\gamma^\us(x',\zeta';x,\zeta)$
from $x$ to $x'$ along $\gamma$ in the arbitrary gauge $\us$ is defined by
\eqn\etprop{K_\gamma^\us(x',\zeta';x,\zeta)=
\bra\uphi_{\zeta'}^{\us(x')}|\,\tau_\gamma(x',x)\,|
\uphi_{\zeta}^{\us(x)}\ket\ ,}
i.e., this is the matrix element of the map $\tau_\gamma(x',x)$, given by
\eptuphiz, between the fibre basis states
$\uphi_{\zeta'}^{\us(x')}\in\pi^{-1}(x')$ and
$\uphi_{\zeta}^{\us(x)}\in\pi^{-1}(x)$, defined in \euphiz. Its definition
is of course exactly analogous to that of the quantum transport propagator
$\uk^\us(x',\zeta';x,\zeta)$ in
\eqprop, and it plays the same role for parallel transport as that
object does for quantum transport. In particular, it too satisfies
equations corresponding to \eqwfprop\ (with $\upsi_{x'}$ defined via
\epltr\ rather than \esecqdef)
%, \eqpcmp\
and \eqprg.

As in the previous section, we now concentrate on the special case where
$(M,g^L)$ is Minkowski space. As parallel transport is path-independent we
will drop the subscript $\gamma$ on $\tau_\gamma(x',x)$ and
$K_\gamma^\us(x',\zeta';x,\zeta)$. In this case there is a remarkable
connection between quantum and parallel transport. Under the affine
connection described by \elsecpm\ and \etilompb, it can be shown that the
section $\us_L$ of $\PM$ defined by \elorsec\ has vanishing covariant
derivative, so that
\eqn\eptsl{\tau(x',x)\,\us_L(x) = \us_L(x')\ .}
Using \eptuphiz\ it follows that
\eqn\eptuphisl{\tau(x',x)\,\uphi_\zeta^{\us_L(x)} =
\uphi_\zeta^{\us_L(x')}\qquad\forall\zeta\in\R^4\times i\V^+\ .}
Now using the formula \eqt\ for the quantum transport map $Q(x',x)$ in the
flat space case, and the definition \euphiz\ of $\uphi_\zeta^{\us}$, we see
that
\eqn\epteqta{\tau(x',x)\,\uphi_\zeta^{\us_L(x)} = Q(x',x)\,
\uphi_\zeta^{\us_L(x)}\ .}
Since the $\uphi_\zeta^{\us_L(x)}$ form a basis for $\pi^{-1}(x)$, it
follows that quantum and parallel transport coincide in the flat space
case,
\eqn\epteqt{\tau(x',x)=Q(x',x)\ .}
The corresponding propagators defined by \eqprop\ and \etprop\ are of
course also equal, in any gauge $\us$,
\eqn\eppeqp{K^\us(x',\zeta';x,\zeta)=\uk^\us(x',\zeta';x,\zeta)\ .}
If the Levi-Civita connection were extended to a connection on $\PM$ having
a non-flat translational part, in distinction to the affine connection
described by \etilompb, then the relation \eppeqp\ would not hold.

As an aside, we remark that combining \eppeqp\ and \eqpqtsl\ gives
\eqn\eqpptsl{K^{\us_L}(x',\zeta';x,\zeta)=K^{(\ell)}(\zeta',\zeta)\ .}
This is in conflict with equation (4.6.5) of reference \rpnewb, which in
the present notation would be
$K^{\us_L}(x',\zeta';x,\zeta)=K^{(\ell)}(x'+\zeta',x+\zeta)$. For example,
\eqpptsl\ implies that $K^{\us_L}(x',\zeta';x,\zeta)$ is independent of
$x'$ and $x$, but the cited equation from reference \rpnewb\ implies an
$x'$ and $x$ dependence\foot{This conflict has recently been
removed in Ref. \rlast.}.

We return briefly to the question of the physical status of the quantum
transport law raised at the end of the previous section. We have shown here
that in the flat space case, quantum transport coincides with parallel
transport. This of course is governed by a first-order differential
equation, the geodesic equation, which does not have the character of a
wave equation enforcing the causality properties of the underlying
spacetime metric. Rather, the causality properties of this description of
the particle are already encoded in any individual state in the fibre above
any point $x\in M$, each such state giving a complete description of the
system. This seems to favour an interpretation of the role of quantum
transport as a {\it kinematical consistency condition\/} rather than as a
dynamical evolution law.

\newsec{Path integral formula for quantum transport in the flat space case}

We will now show that the quantum propagator $\uk^\us(x',\zeta';x,\zeta)$
in flat space satisfies an equation which is similar to \eqpcmp, but where
the integral involves a spacelike surface in the base spacetime, rather
than being an integral over a $\zeta$ variable for a fixed point in the
base spacetime.  Combining \eqpcmp\ and \eqprg\ we find
\eqn\epia{\eqalign{\uk^\us(x',\zeta';x,\zeta) =
\int_{\Sigma^+}\uk^{\us_L}(&x',g_\us(x')^{-1}\zeta';x_1,g_\us(x_1)^{-1}
\zeta_1)\cr&
\uk^{\us_L}(x_1,g_\us(x_1)^{-1}\zeta_1;x,g_\us(x)^{-1}\zeta)
\,d\Sigma(\zeta_1)\ ,\cr}}
where $\us$ is an arbitrary Poincar\'e gauge, $\us_L$ is defined by
\elorsec, and $g_\us$ is defined by \egs. Using the Poincar\'e invariance
of the measure $d\Sigma$ (equivalently the unitarity of the representation
\eprep) we can eliminate $g_\us(x_1)$, giving
\eqn\epib{\uk^\us(x',\zeta';x,\zeta) =
\int_{\Sigma^+}\uk^{\us_L}(x',g_\us(x')^{-1}\zeta';x_1,\zeta_1)
\,\uk^{\us_L}(x_1,\zeta_1;x,g_\us(x)^{-1}\zeta)\,d\Sigma(\zeta_1).}
Now we use the fact, noted after equation \eqpqtsl, that
$\uk^{\us_L}(x',\zeta';x,\zeta)$ is independent of $x'$ and $x$, writing
\epib\ in terms of $K^{(\ell)}$, to allow us to relabel $q_1$ in
$\zeta_1=q_1-i\ell v_1$ as $x_1$, which in the gauge $s_L$ is identical to
$-a_L(x_1)$. (Recall that $x_1=(x_1^\mm)\in\R^4$ is the vector of
coordinates of a point in $M$, rather than being the abstract vector $x$
%$\ux\in M$ ($\ux=x^\mm\ue_\mm(O)$)
appearing in footnote \fxbar, so the equation $x_1=q_1$ is well-defined.)
This gives
\eqn\epic{\eqalign{\uk^\us(x',\zeta';x,\zeta)=
\int_{\Sigma^+}\uk^{\us_L}(x'&,g_\us(x')^{-1}\zeta';x_1,x_1-i\ell v_1)
\cr\uk^{\us_L}&(x_1,x_1-i\ell v_1;x,g_\us(x)^{-1}\zeta)
\,d\Sigma(x_1-i\ell v_1)\ .\cr}}
Thus the integral on the right hand side has been recast in a form such
that the $x_1$ integral can be thought of as going over a spacelike surface
$\sigma$ {\it in the base Minkowski space-time\/} (where
$\Sigma^+=\sigma\times\V^+$, see \esig), albeit that this arose rather
trivially from the fact that the integrand in \epib\ was independent of
$x_1$. Using \eqprg\ again, we write the integral back in terms of
propagators in the original gauge $\us$,
\eqn\epid{\eqalign{\uk^\us(x',\zeta';x,\zeta)=
\int_{\Sigma^+}\uk^{\us}(x'&,\zeta';x_1,g_\us(x_1)(x_1-i\ell v_1))\cr
\uk^{\us}&(x_1,g_\us(x_1)(x_1-i\ell v_1);x,\zeta)
\,d\Sigma(x_1-i\ell v_1)\ .\cr}}
(The Poincar\'e invariance of the measure $d\Sigma$ cannot be used to
eliminate $g_\us(x_1)$ from this expression. The obvious substitution
$x_1'-i\ell v_1' = g_\us(x_1)(x_1-i\ell v_1)$ of course eliminates it
from the arguments in which it currently appears, but it reappears upon
substituting the two arguments which involve $x_1$ alone, since $x_1 =
g_\us(x_1)^{-1}x_1'$ (see \epgq, \epgzeta).) Now we simplify the
expression $g_\us(x)(x-i\ell v)$. Let the arbitrary gauge $\us$ be
$\us(x)=(\ua(x),\ue_\mm(x))$, as in
\esecpm, and let $g_\us(x) = (b(x),\Lambda(x))$. {}From \epgzeta\ we have,
since $x=-a_L(x)$ in the gauge $s_L$ according to \elorsec,
%\eqn\epie{g_\us(x)^{-1}(x-i\ell v)=
%\Lambda(x)^{-1}(x-b(x)) -i\ell\Lambda(x)^{-1}v\ .}
\eqn\epie{g_\us(x)(x-i\ell v)=
%g_\us(-a_L(x)-i\ell v)=
-\Lambda(x)a_L(x) - b(x) -i\ell\Lambda(x)v\ .}
%Let $a(x)=(a^\mm(x))$, where $a^\mm(x)$ are the components of $\ua(x)$,
%defined as in \eacpt.
%Recall that in the gauge $\us_L$ defined by \elorsec\
%the components of $\ua_L(x)$ are $a_L(x)=x$.
Then from \egs\ and the
definition \era\ of the right action of $\pg$ on $\PM$, we find that
%{}From \era\ and \egs\ we find
%\eqn\epif{a(x) = \Lambda(x)^{-1}(b(x)-x)\ ,}
\eqn\epif{a(x) = \Lambda(x)a_L(x) + b(x)\ .}
%where $a(x)=(a^\mm(x))$, and the $a^\mm(x)$ are defined as in
%\eacpt.
We emphasise that this formula for $a(x)$ holds for each $x$, and merely
displays the relationship between the components of $\ua(x)$ corresponding
to the arbitrary gauge $\us$, and the components of $\ua_L(x)$
corresponding to the gauge $\us_L$, which relationship is given at each
point $x$ by the Poincar\'e transformation $g_\us(x) = (b(x),\Lambda(x))$.
Combining these results and writing it in terms of vectors of components
$a^i(x)$ and $v^i$ respectively gives
%\eqn\epig{g_\us(x)^{-1}(x-i\ell v) = -a(x)-i\ell\Lambda(x)^{-1}v\ .}
%\eqn\epig{g_\us(x)(x-i\ell v) = -a(x)-i\ell\Lambda(x)v\ .}
\eqn\epig{g_\us(x)(x-i\ell v) = -a(x)-i\ell\bar v(x)\ ,}
where $\bar v(x)=\Lambda(x)v$.  Here the vector $-a(x)$ denotes the point
of contact of $T_xM$ and $M$ in the gauge $s$ in $T_xM$ with $M$ being
Minkowski space in the present case. Furthermore, $\Lambda(x)$ relating $v$
and $\bar v$ in \epig\ is the Lorentz transformation by which the special
``parallel'' gauge $s_L$ deviates from the general Poincar\'e gauge $s$.

Substitution in \epid\ now gives
%\eqn\epih{\eqalign{\uk^\us(x',\zeta';x,\zeta)=
%\int_{\Sigma^+}\uk^{\us}(x',\zeta';&x_1,-a(x_1)-i\ell\Lambda(x_1)^{-1}v_1)\cr
%\uk^{\us}(&x_1,-a(x_1)-i\ell\Lambda(x_1)^{-1}v_1;x,\zeta)
%\,d\Sigma(x_1-i\ell v_1)\ .\cr}}
\eqn\epih{\eqalign{\uk^\us(x',\zeta';x,\zeta)=
%\int_{\Sigma^+}\uk^{\us}(x'&,\zeta';x_1,-a(x_1)-i\ell\Lambda(x_1)v_1)\cr
\int_{\Sigma^+}\uk^{\us}(x'&,\zeta';x_1,-a(x_1)-i\ell\bar v(x_1))\cr
%\uk^{\us}&(x_1,-a(x_1)-i\ell\Lambda(x_1)v_1;x,\zeta)
\uk^{\us}&(x_1,-a(x_1)-i\ell\bar v(x_1);x,\zeta)
%\,d\Sigma(x_1-i\ell v_1)\ .\cr}}
\,d\Sigma(-a(x_1)-i\ell\bar v(x_1))\ ,\cr}}
where we have used \epig\ and the Poincar\'e invariance of the measure to
put it into the form $d\Sigma(-a(x_1)-i\ell\bar v(x_1))=d\Sigma(x_1-i\ell
v_1)$.
Finally, this expression can be substituted into itself an arbitrary number
of times to obtain
\eqn\epi{\eqalign{\uk^\us(x',\zeta';x,\zeta)=
\int_{\Sigma^+_1}\ldots&\int_{\Sigma^+_{N-1}}
\uk^\us(x',\zeta';x_{N-1},\hat\zeta_{N-1})\cr&
\prod_{n=N-1}^1\uk^\us(x_n,\hat\zeta_n;x_{n-1},\hat\zeta_{n-1})\,d
%\Sigma(x_n-i\ell v_n)\ ,\cr}}
\Sigma(-a(x_n)-i\ell\bar v_n)\ ,\cr}}
where
%\eqn\epii{\hat\zeta_n = -a(x_n) - i\ell\Lambda(x_n)^{-1}v_n\ ,
%\qquad n=1,\ldots,N-1\ ,}
\eqn\epii{
%\hat\zeta_n = -a(x_n) - i\ell\Lambda(x_n)v_n\ ,
\hat\zeta_n = -a(x_n) - i\ell\bar v_n\ ,
\qquad n=1,\ldots,N-1\ ,}
and
\eqn\epij{x_0 = x\ ,\qquad \hat\zeta_0 = \zeta\ ,}
for any integer $N\ge1$.

In reference \rpnewb\ equations (4.6.7a,b), \prug\ postulates a Feynman
path integral-like equation for the quantum transport propagator in the
Hilbert bundle over a curved base spacetime. The flat space limit of this
equation should of course be consistent with our equation \epi\ just
derived. We now examine this consistency. The use of the parallel transport
propagator $K^\us$ rather than $\uk^\us$ in the integrand of \rpnewb\
eq.(4.6.7a) is of course justified, since these coincide in the flat space
case (equation \eppeqp). Further, each of the surfaces $\Sigma^+_n$,
$n=1,\ldots,N-1$, in \epi\ is of the form $\sigma_n\times\V^+$, where
$\sigma_n$ is an arbitrary Cauchy surface in Minkowski space (see equation
\esig). One possible choice would be to take these surfaces to be
time-ordered, discrete leaves in a foliation of a region of spacetime
bounded by two (non-intersecting) spacelike hypersurfaces $\sigma_0$ and
$\sigma_N$, with $\sigma_0$ containing $x$ and $\sigma_N$ containing
$x'$. Then considering the limit $N\rightarrow\infty$ would indeed produce
a path integral-like formula of the form of \rpnewb\ eq.(4.6.7a) provided
the measure $d\Sigma(x_n,\bar v_n)$ appearing there is identified in flat
space with the measure $d\Sigma(-a(x_n)-i\ell\bar v_n)$ on the right
hand side of \epi\ where $-a(x_n)$ denotes, as mentioned, the coordinates
of the point $x\in M\equiv\R^4$with regard to the affine base in $T_xM$ in
the gauge $s$.

\newsec{Possible extensions of quantum transport to curved spacetime}

The approach of reference \rpnewb\ to postulating a definition for the
quantum transport propagator in the Hilbert bundle over a curved
spacetime base is motivated by the form of this propagator in the flat space
case. The flat space quantum propagator can be written in a way involving
the parallel transport propagator which, it is argued, has a natural
extension to curved space, obtained simply by replacing the flat space
parallel transport propagator by the one for curved space. However, this
generalisation to curved spacetime is not immediate. The
difficulty is the non-existence in the case of curved spacetime of a unique
gauge choice analogous to $\us_L$ in the flat space case. This is because
the path-dependence of parallel transport makes it impossible to construct
a Lorentz frame at each point, such that the parallel transport of the
frame from one point to any other point gives the frame at that point,
independent of the path taken.

There is one further point to be considered. In generalizing the expression
for the quantum propagator from a flat spacetime base to a curved spacetime
base the hypersurfaces $\Sigma^+=\sigma\times\V^+$ defined in \esig, with
$\sigma$ being a space-like surface in Minkowski space, has to be turned
into a spacelike hypersurface in curved spacetime representing a leaf of a
foliation of the underlying curved manifold. It is not clear from the
beginning whether the invariant measure \efmeas\ defining our Hilbert space
$\F$, which carries a unitary irreducible representation of the Poincar\'e
group of spin $s=0$ and mass $m$, is immediately interpretable as a measure
on the different hypersurfaces $\sigma_n\times\V^+$ where the $\sigma_n$
are given by a particular foliation of the curved spacetime base, and
whether the whole procedure of defining a quantum propagator on curved
spacetime is strongly dependent on the foliation chosen.

Let us now generalize eq.\epi\ in a way which is in accord with \prug's
proposal presented in \rpnewb\ eq.(4.6.7a,b) yielding, indeed, a Poincar\'e
gauge covariant definition (see
eq.\eqprg) of a quantum-geometric propagator defined on a one-particle
(first quantized) Hilbert bundle over curved spacetime which is based on
parallel transport along geodesic paths denoted by $\gamma(x_n,x_{n-1})$
between points $x_{n-1}\in\sigma_{n-1}$ and $x_n\in\sigma_n$ of two
adjacent leaves of a foliation\foot{In \rpnewb\ the geodesic arc between
$x_{n-1}$ and $x_n$ was denoted by $\gamma(x_{n-1},x_n)$.}. We show that
the resulting Feynman-like path integral expression involving integrations
over intermediate points on $\sigma_n\times\V^+$, with the $\sigma_n$
being, as mentioned, hypersurfaces of a foliation in the spacetime base
rather that hypersurfaces in the fibre, is a Poincar\'e gauge covariant
expression; or, more precisely, obeys the relation
\eqn\eqprgt{
\uk^{\us'}(x',\zeta';x,\zeta)=
\uk^{\us}(x',g(x')^{-1}\zeta';x,g(x)^{-1}\zeta)}
with $g(x)=(b(x),\Lambda(x))$ and $s$ and $s'$ related as in \egdef. In
accord with \rpnewb\ eq.(4.6.7a,b) we thus write in the gauge $s$ for the
quantum propagator on a curved base manifold $M$
\eqn\eprprop{\eqalign{
\underline K^{s}(x',\zeta';x,\zeta)
={\hbox{lim}\atop\epsilon\rightarrow+0}
\int &K^{s}_{\gamma(x',x_{N-1})}(x',\zeta';x_{N-1},\hat\zeta_{N-1})\cr&
\prod_{n=N-1}^1 K^{s}_{\gamma(x_n,x_{n-1})}(x_n,\hat\zeta_n;x_{n-1},
\hat\zeta_{n-1})\,d\Sigma(x_n,\bar v_n)\ ,}}
with $\epsilon=(t'-t)/N$ and $\hat\zeta_n$ given by
\eqn\eprpropa{\hat\zeta_n=-a(x_n)-i\ell\bar v_n\ ,
\quad n=1,\ldots,N-1\ ,\qquad
%}
%and
%\eqn\eprpropb{
x_0 = x\ ,\quad\hat\zeta_0 = \zeta\ ,}
and where we have introduced a measure $d\Sigma(x_n,\bar v_n)$ to be
defined below. In \eprprop\ the parallel transport propagator between two
points on adjacent leaves of a foliation connected by a geodesic arc
$\gamma(x_n,x_{n-1})$ from $x_{n-1}$ to $x_n$ is denoted by
$K^{s}_{\gamma(x_n,x_{n-1})}(x_n,\hat\zeta_n;x_{n-1},
\hat\zeta_{n-1})$ where the suffix $\gamma$ is a reminder of the path
dependence. Eq.\era\ implies that two gauges $s$ and $s'$ are related by
\eqn\etgs{\eqalign{&s(x)=(\ua(x)=-a^\mm(x)e_\mm(x),e_\mm(x))\ ,
\qquad s'(x)=(\ua'(x)=-a'^\mm(x)e'_\mm(x),e'_\mm(x))\cr
&\hbox{with}\qquad
a'^\mm(x)=[\Lambda(x)]^\mm{}_\nn a^\mm(x)+b^\mm(x)
\qquad\hbox{and}\qquad e'_\nn(x)=e_\nn(x)[\Lambda(x)^{-1}]^\mm{}_\nn\cr}}
and the corresponding $\zeta$ variables defined with respect to the frames
$s(x)$ and $s'(x)$ transforming as
\eqn\etgsb{\zeta'(x)=g(x)\zeta(x)=\Lambda(x)q(x)-b(x)-i\ell\Lambda(x)\bar
v(x)\ .}
We mention again that the point $q(x)=-a(x)$ is the coordinate description
of the point of contact $x\in M$ as determined in the gauge $s$ in $T_xM$
and analogously in the gauge $s'$.

Let us now use \eqprgt\ as well as \etgs\ and \etgsb\ to convert \eprprop\
into an expression for $\underline K^{s'}$. We obtain (with $x_0=x$ and
$g(x_0)\hat\zeta_0=\zeta$)
\eqn\eprpropt{\eqalign{
\underline K^{s'}(x',\zeta';x,\zeta)=\qquad&\cr
{\hbox{lim}\atop\epsilon\rightarrow+0}
\int &K^{s'}_{\gamma(x',x_{N-1})}(x',\zeta';x_{N-1},g(x_{N-1})
\hat\zeta_{N-1})\cr&
\prod_{n=N-1}^1 K^{s'}_{\gamma(x_n,x_{n-1})}(x_n,g(x_n)
\hat\zeta_n;x_{n-1},
g(x_{n-1})\hat\zeta_{n-1})\,d\Sigma(x_n,\bar v_n)\ .}}
Now we need to evaluate $g(x_n)\hat\zeta_n$ for the intermediate points
which is, for $g(x_n)=(b(x_n),\Lambda(x_n))$ using \eprpropa\ and \epgzeta,
given by
\eqn\eintpt{g(x_n)\hat\zeta_n=-\Lambda(x_n)a(x_n)-b(x_n)-i\ell\Lambda(x_n)v_n
= -a'(x_n)-i\ell\bar v'_n = \hat\zeta'_n\ .}
According to \etgs\ this is the gauge transformed variable
$\hat\zeta'_n=\zeta'(x_n)$ with $\bar v'_n=\Lambda(x_n)\bar v_n$.
It is now seen from \etgsb\ together with \eintpt\ that the quantum
propagator for a curved base manifold $M$ is definable in a Poincar\'e
gauge invariant manner provided the measure $d\Sigma(x_n,\bar v_n)$ is
Poincar\'e invariant.

The original measure $d\Sigma(\zeta)=d\Sigma(q,v)$ defined in \efmeas\ is
Poincar\'e invariant. Clearly, the measure denoted by $d\Sigma(x_n,\bar
v_n)$ in \eprprop\ and \eprpropt\ cannot be invariant against arbitrary
reparametrizations of the base manifold given in terms of changes of the
atlas on $M$. The only way the notation $d\Sigma(x_n,\bar v_n)$ for the
measure makes sense is by interpreting $x_n$ as the {\it point of contact}
of $T_xM$ with the base $M$ as given in terms of the coordinates of this
point in the gauge $s$ {\it in the local tangent space\/} $T_xM$, which is
given by $-a(x_n)$, i.e., by interpreting $d\Sigma(x_n,\bar v_n)$ as
$d\Sigma(-a(x_n),\bar v_n)=d\Sigma(\hat\zeta_n)$\foot{Despite the complex
notation adopted for $\zeta=q-i\ell v$ following \rpnewb\ we may denote the
measure $d\Sigma(\zeta)$ equivalently by $d\Sigma(q,v)$.}.
Of course, there does not
exist now a gauge $s_L$ as in the Minkowski case, described in sections 4
and 6 above, where $x$ is equal to $-a_L(x)$ for all
$x\in\R^4$. Interpreting the measure appearing in \eprpropt\ in this manner
as a ``contact point measure'' one can show that this measure is Poincar\'e
gauge invariant since then one has using
$d\Sigma(\zeta)=d\Sigma(g(x)\zeta)$:
\eqn\emeast{\eqalign{d\Sigma(x_n,\bar v_n)&\equiv d\Sigma(-a(x_n),\bar v_n)
=d\Sigma(g(x_n)(-a(x_n)),g(x_n)\bar v_n)\cr&
=d\Sigma(-a'(x_n),\bar v'_n)\equiv d\Sigma(x_n,\bar v'_n)\cr}}
with the primed quantities given as in \eintpt.

Hence we come to the conclusion that \eprprop\ can, with the help of
\eintpt\ and \emeast, be written in terms of the primed variables belonging
to the gauge $s'$ showing that \eprprop\ is a Poincar\'e gauge covariant
quantum-geometric propagator defined on a first quantized Hilbert bundle
over curved spacetime with the measure $d\Sigma(x_n,\bar v_n)$ interpreted
as a ``contact point measure'' in the above described sense. Clearly,
eq.\eprprop\ is a particular proposal for a Poincar\'e gauge invariant
quantum propagator on a curved base which is motivated by the flat space
result investigated in section 6 depending, moreover, on the particular
foliation chosen for the spacetime base.

In concluding this section we would like to add one further remark. In
order to interpret the ``contact point measure'' $d\Sigma(x_n,\bar v_n)$ as
a bona fide measure definable on a general curved spacetime base one would
have to present an analytic proof that it is possible to go over in a
smooth manner on $M$, from a truncated measure defined on the local fibres
of the phase space bundle
\eqn\eppb{\tilde E=\tilde E(M,F=\R^4\times\V^+,\pg)\ ,}
soldered to the base $M$ through the subspace $\R^4$ of $F$ -- i.e., from a
measure restricted to the point of contact of fibre and base in $\tilde E$
-- to a smooth Poincar\'e invariant measure defined on the base with values
in $\V_x^+$. (Compare also footnotes \fftxmvx\ and \ffxp\ above in this
context.) The soldering maps could be used in building up a hypersurface
$\sigma$ in the spacetime base from the surface $\Sigma^+_x$ defined in the
local fibres of $\tilde E$ using analyticity in the affine variable $q\in
T_xM$ in order to construct a Cauchy surface in $M$ generating a
foliation. To our knowledge, investigations in this direction have not been
made.

\newsec{Modifications to incorporate the usual description of the
Klein-Gordon particle}

The Hilbert bundle description can equally well be applied in the context
of the conventional description of relativistic quantum mechanics, without
introducing the full stochastic phase-space quantum mechanics (and thereby
of course foregoing its advantages). We briefly sketch this here.

The conventional, configuration-space description of the Klein-Gordon
particle is obtained by applying the Fourier transform to the
momentum-space wavefunctions, instead of equation \efdef. That is, instead
of $\F$, the Hilbert space is $\F'$, consisting of the states $\psi$ on
$\R^4$ of the form
%\eqn\enkgwf{\psi(q)= 2(2\pi)^{-3/2}\int_{\V_m^+} e^{-iq.k}
%\tilde\psi(k)d\Omega_m(k)\ .}
\eqn\enkgwf{\psi(q)= (2\pi)^{-3/2}\int_{\V_m^+} e^{-iq.k}
\tilde\psi(k)d\Omega_m(k)\ ,}
where $\tilde\psi\in L^2(\V_m^+)$. $\F'$ carries a spin zero, irreducible
representation of the Poincar\'e group, and a Hilbert bundle with typical
fibre $\F'$, associated to the Poincar\'e frame bundle, can be constructed
as before, including the analogues of the canonical maps $\usig_x^\us$
(equation \ecnmap). In this case, there is no (covariant) configuration
space probability interpretation for the states $\psi$ in
$\F'$. Nevertheless, one can make a connection between the bundle and
non-bundle formulations in the same fashion as before, by defining the
configuration space amplitude corresponding to a section of the (new)
Hilbert bundle via equation \eprob, with the $\uv$ variable
suppressed. Despite the absence of a physical (probability) interpretation
for such amplitudes, it is natural to require each such amplitude to agree
with some wavefunction in $\F'$. Perhaps the strongest argument for this is
that we are effectively considering the one particle sector of the second
quantized theory, and agreement of the analogous amplitudes in the
non-bundle and bundle descriptions of the second quantized theory is
required for physical reasons. The succeeding analysis proceeds as before,
with the change that the momentum variables are everywhere suppressed.

\newsec{Conclusion}

We have examined in detail the Hilbert bundle formulation of the
relativistic phase space quantum theory of a spinless particle. In the flat
space limit we deduced the required quantum transport law and propagator on
the bundle by comparison with \prug's stochastic phase space quantum
mechanics. We showed that this coincides with parallel transport on the
bundle (in the flat space limit). We found furthermore, that a formula for
the quantum propagator in curved spacetime, which was postulated in \rpnewb,
is Poincar\'e gauge covariant (as stated in \rpnewb) provided the measure
is interpreted as a ``contact point measure'' on the soldered bundle
$\tilde E$. The problems associated with this interpretation were briefly
mentioned. Finally, we sketched briefly how the Hilbert bundle construction
can also be applied to the conventional description of the Klein-Gordon
particle.

It is apparent that the Hilbert bundle approach gives an elegant and powerful
new
framework for quantum field theory in curved spacetime which has
fundamental advantages over the conventional approach. We believe that further
useful work can be done on the question of specifying the integration measure
for the quantum propagation law in curved spacetime,
for example along the lines mentioned in section 7 above.

\noindent{\bf Acknowledgements}\par\nobreak\medskip\nobreak

We are grateful to Professor \prug\ for sending us advance copies of parts
of reference \rpnewb\ and for discussions and correspondence in the final
stage of writing this paper. We are also grateful for receiving
prior to publication a copy of reference \rlast\ treating similar problems to
those
discussed here.

\listrefs

\bye